\documentclass[%
 aip,
cp,  
 amsmath,amssymb,
 reprint,%
]{revtex4-2}

\usepackage{graphicx}
\usepackage{dcolumn}
\usepackage{bm}
\usepackage{wrapfig}

\usepackage[utf8]{inputenc}
\usepackage[T1]{fontenc}
\usepackage{mathptmx} 
\usepackage{natbib}
\usepackage[english]{babel}
\usepackage{setspace}
\usepackage[utf8]{inputenc}
\usepackage{polski}
\usepackage[T1]{fontenc}
\usepackage{graphicx}
\usepackage{wrapfig}
\usepackage{txfonts}
\usepackage{amsmath}
\usepackage{float}
\usepackage{natbib}
\usepackage{graphicx}
\usepackage{txfonts}
\usepackage{amsmath}
\usepackage{float}
 \usepackage[table,xcdraw]{xcolor}
\usepackage[hidelinks,colorlinks=true,linkcolor=blue,citecolor=blue]{hyperref}
\begin{document}

\title{Transformers as Strong Lens Detectors- From Simulation to Surveys}

\author{Hareesh Thuruthipilly} 
 \email[Corresponding author: ]{hareesh.thuruthipilly@ncbj.gov.pl}
\author{Margherita Grespan}%
 \email[Corresponding author: ]{margherita.grespan@ncbj.gov.pl}
 \author{ Adam Zadrożny}%
\affiliation{National Centre for Nuclear Research, Astrophysics division (BP4),
              ul. Pasteura 7, 02-093 Warszawa.}


\date{\today} 

\begin{abstract}
With the upcoming large-scale surveys like LSST, we expect to find approximately $10^5$ strong gravitational lenses among data of many orders of magnitude larger. In this scenario, the usage of non-automated techniques is too time-consuming and hence impractical for science. For this reason, machine learning techniques started becoming an alternative to previous methods. In our previous work \citep{Hareesh}, we have proposed a new machine learning architecture based on the principle of self-attention, trained to find strong gravitational lenses on simulated data from the Bologna Lens Challenge. Self-attention-based models have clear advantages compared to simpler CNNs and highly competing performance in comparison to the current state-of-art CNN models. We apply the proposed model to the Kilo Degree Survey, identifying some new strong lens candidates. However, these have been identified among a plethora of false positives, which made the application of this model not so advantageous. Therefore, throughout this paper, we investigate the pitfalls of this approach, and possible solutions, such as transfer learning, are proposed.
\end{abstract}

\maketitle

\section{\label{sec:level1} Introduction}

Strong gravitational lensing is one of the interesting predictions of General Relativity (GR); it describes how light rays from a distant astronomical source are deflected due to the massive gravitational potential of a foreground galaxy. The manifestation of this effect in nature is observed as arcs or rings around massive galaxies or multiple images of the source. Different lens-source configurations produce different distortions. For a detailed review, see \citep{Treu_2010}.  Strong lensing (SL) systems are employed for probing many unique astrophysical and cosmological phenomena. For instance, they act as "gravitational telescopes", enabling the study of distant astronomical objects which would be otherwise too faint to observe, such as high redshift galaxies \citep{Ebeling, Richard}, dwarf galaxies \citep{Marshall}, etc. Depending on the source and the lens configuration, there are various kinds of strong lensing systems, such as galaxy clusters acting as lenses, quasars or supernovas acting as sources, etc. One particular case of gravitational lensing is galaxy-galaxy strong lensing (GGSL), where the background source and foreground lens are both galaxies and massive enough to produce distortion in the images of the source.

Ever since the discovery of the first GGSL system by \citet{Hewitt}, numerous applications of GGSL in cosmology and astrophysics have been discovered \citep{Koopmans_2006,2009ApJ...691..531C, Collett_2014, Cao_2015, Bonvin_2016}. Strong lensing has been used to derive cosmological constraints on dark energy \citep{Shiralilou} and on the cosmic expansion history \citep{Biesiada2006, Jullo, Caminha}. Strong lensing can also be used to constrain dark matter models \citep{Vegetti, Diaz, Diaz2018, Nadler}, as well as to detect dark-matter substructures \citep{Despali}. For these reasons, the current and upcoming surveys have significantly focused on detecting SLs. For a detailed review of the applications of strong lensing, please refer to \citet{1992ARA&A..30..311B}.

However, as promising and exciting as it sounds, strong gravitational lensing is a rare phenomenon, with an order of magnitude of one gravitational lens per thousand galaxies observed \citep{Chae_2002,Wardlow_2013}. For all the applications mentioned above, a large sample of SLs is required to obtain statistically significant results. Unfortunately, only a few hundred lensing systems have been detected and confirmed by the present astronomical surveys until now. With the upcoming era of advanced missions such as the Euclid \citep{scaramella2021euclid}, and LSST \citep{Ivezi__2019,verma2019strong}, the number of observable SLs is expected to reach $10^5$ \citep{Oguri_2010,Collett_2015,McKean_2015}, which should be identified among $10^9$ galaxies \citep{Wilde_2022}. Similarly, the number of new SLs expected to be discovered in the Square Kilometre Array (SKA) survey will also have similar orders of magnitude \citep{2015aska.confE..84M}. Therefore, analyzing data on future large-scale surveys with  manual or semi-automatic methods is highly time-consuming and impractical. Hence, the situation demands better and more effective alternative approaches, which should be fast, reliable, and robust.

In recent times, advancements in deep learning and computer vision have found applications in astronomy. In particular, Convolutional Neural Networks (CNNs) have been able to emerge as the state-of-the-art algorithm in various sectors of astronomy (e.g. galaxy classification by \citealt{2019PASP..131j8002P}, supernova classification by \citealt{2017ApJ...836...97C}, and galaxy merger identification by \citealt{Louis}). Similarly, CNNs and other deep learning-based algorithms have been found to be effective in detecting and analyzing SLs from large-scale surveys \citep{ Lanusse_2017, Schaefer_2018,2019MNRAS.487.5263D,2020MNRAS.496..381C}. 
For instance, \citet{2017MNRAS.471..167J} applied CNNs to the data from Canada-France-Hawaii Telescope Legacy Survey (CFHTLS) to find SLs. Similarly, numerous other successful attempts to find potential SL candidates from the Kilo Degree Survey (KiDS) have been reported \citep{2017MNRAS.472.1129P, Petrillo_2018, Petrillo_2019, He_2020, Li_2020}. Furthermore, various groups have successfully used CNNs to identify strong lens galaxy-scale systems from large-scale surveys such as Dark Energy Survey (DES) \citep{2019ApJS..243...17J, rojas2021strong}, Dark Energy Spectroscopic Instrument Legacy Imaging Surveys  \citep{Huang_2020, Huang_2021}, Pan-STARRS \citep{Lens_CNN&A...644A.163C} and also on comparatively small scale surveys like VOICE \citep{2021arXiv210505602G}. Furthermore, the Strong Gravitational Lens Finding Challenge, designed to compare and develop new lens-finding approaches, has demonstrated that CNNs perform better than human inspection or any other traditional methods \citep{Metcalf_2019}.

Recently, there was a breakthrough in natural language processing (NLP) by \citet{vaswani2017attention} with the introduction of self-attention-based architecture known as the transformers. Since then, there have been attempts to adapt the idea of self-attention to build better image processing models \citep{ramachandran2019standalone,zhao2020exploring,tan2021explicitly}. The basic idea behind the transformer architecture is the attention mechanism, which has also found a wide variety of applications in machine learning \citep{zhang2019selfattention,fu2019dual}. In the case of NLP, self-attention correlates the different positions of a single sequence to calculate a representation of the sequence.
Similarly, the idea of self-attention, as the name suggests,  is to give relative significance to the input features based on the input features themselves, which helps the network to create a representation of the input with the relatively essential features only. Recently, Facebook Inc. \citep{carion2020endtoend} and Google Brain \citep{dosovitskiy2020image} have been able to surpass the existing image recognition models with transformer-based architectures. 

Even though transformer models have overtaken the image processing sector in the machine learning regime,  it is still an unexplored area in astrophysics. Hence, to investigate the potential capacity of attention-based models to find strong lenses, we have implemented various self-attention-based encoder models (transformer encoders) to find the gravitational lenses in the Bologna Lens Challenge \citep{Hareesh}. In our earlier work, we also compared the performance of the encoder models to high-accuracy CNNs taking part in the challenge \citep{Hareesh}. Attention-based models showed high reliability for identifying SL, outperforming most of the CNNs of the Bologna lens challenge. Since the testing on the simulated dataset was a success, we did a blind search (no data pre-processing) of lenses on the Kilo Degree Survey (KiDS) data release 4 (DR4). 
Testing the model on the KiDS survey showed the difficulties of moving from simulated data to real data, showing the limited ability of the model to generalize SL detection. The training data are simulated with respect to the KiDS survey, which makes it possible to apply the model trained on simulations to the survey data. The application of the model gives some candidates which need a visual inspection by human experts further confirm the prediction.  

The paper is organized as follows. Section 2 briefly describes the data used to train and test our models. Section 3 provides a brief overview of the methodology used in our study, including the model's architecture and the procedure we adapted for human visual inspection. The results of our analysis and a detailed discussion of our results are presented in sections 4 and 5, respectively. Finally, we conclude in section 6.

\section{\label{sec:data} Data}

\subsection{\label{sec:Simulated lenses} Simulated lenses - Bologna Lens Challenge}

\begin{figure}[ht]
\centering
\includegraphics[width=480 pt,keepaspectratio]{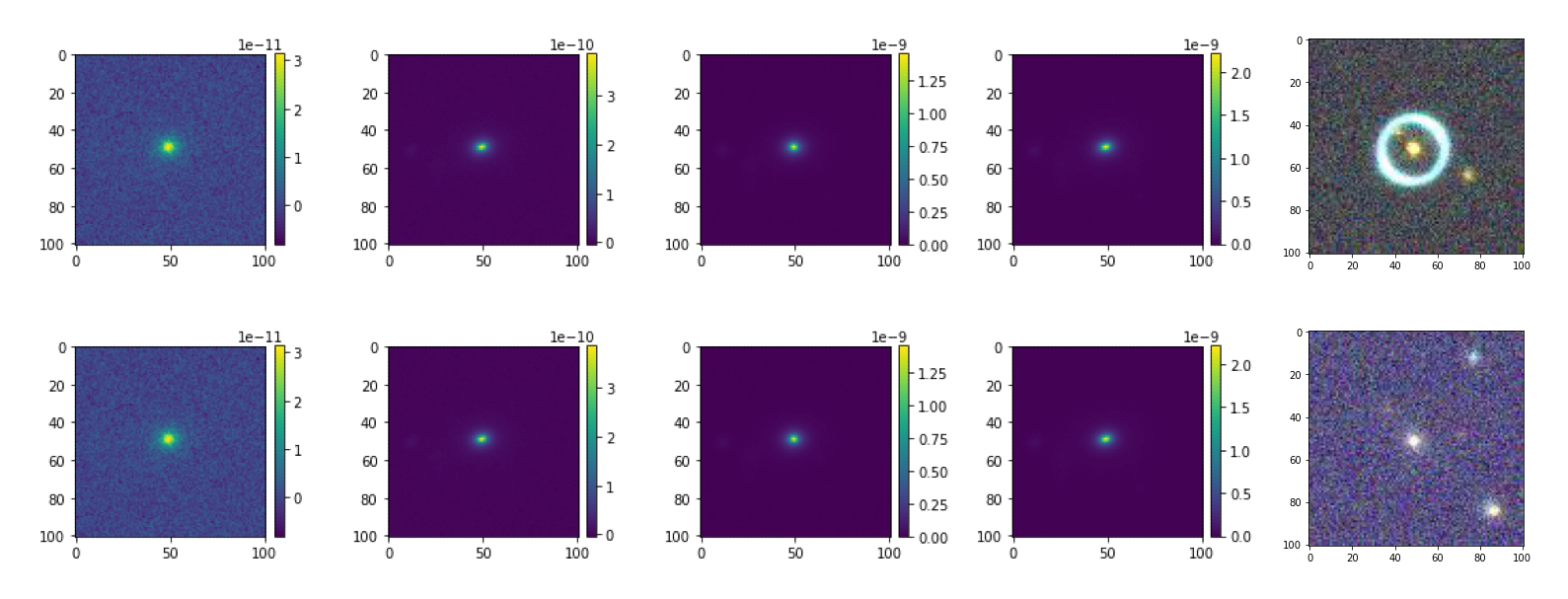}
\caption{Typical image of a mock simulated lens (above) and a non-lens (below) for the challenge. Bands are shown in the following order: $u,g,r,i$, and the corresponding RGB image.}
\label{fig:lenses}
\end{figure}

The novel model for SL detection presented in \citep{Hareesh} is trained on the Bologna Strong Gravitational Lens Finding Challenge \citep{Metcalf_2019}. This challenge required participants to classify strong lenses among other kinds of sources. The images in this dataset are mock observations based on the Kilo-Degree Survey (KiDS).
However, the simulated images do not strictly mimic the surveys; they are only employed as references to set noise levels, pixel sizes, sensitivities, and other parameters. During the challenge, machine learning-based methods were able to classify the images with high accuracy where a human would have doubt. The mock images for the challenge are created using Millennium simulation and GLAMER lensing code \citep{2014MNRAS.445.1942M, Metcalf}. The ground-based images consist of simulated images from four bands $(u, g, r,$ and $i)$, and the reference band was the $r$ band. The noise for the mock images is simulated by adding normally distributed numbers with the variance given by the weight maps from the  KiDS survey. The example images of a mock simulated lens and non-lens for the challenge are shown in Fig. \ref{fig:lenses}. For a detailed review of how the data was   created, please refer to \citet{Metcalf_2019}.

\subsection{\label{sec:Real_Data} Real Data - Kilo-Degree Survey}
The Kilo-Degree Survey \citep{dejong2013} is an ESO
public large optical imaging survey of the Southern sky. It is carried out with the OmegaCAM wide-field imager \cite{Kuijken2011} mounted on the VLT Survey Telescope \citep{capaccioli2011vlt} at the Paranal Observatory in Chile.
The OmegaCAM has a field of view of 1 deg$^2$, and the angular scale of the pixels is 0.21 arcsec. KiDS has surveyed $\approx$ 1350 deg$^2$ of sky in the four optical bands \textit{u,g,r,i} of which 1006 tiles, corresponding to $\approx$ 1 deg$^2$,  are publicly available (data release 4, DR4 \citep{Kuijken_2019}).
The survey has the best seeing in the $r$-band,  with a median point spread function (PSF) full width at half-maximum (FWHM) values of 1.0, 0.8, 0.65, and 0.85 arcsec in
the \textit{u, g, r, and i} bands, respectively.
For this study, from the complete KiDS DR4 (1006 tiles of $\sim$1 deg$^2$), we randomly select $\sim$ 200 tiles as an initial test. For these selected tiles we further sub-sample with a redshift cut for the lens, $z_L$<0.8, giving an average amount of elements per tile of $\approx$23000.  
The KiDS collaboration carried out a search for strong lenses with a convolutional neural network finding 169 SL candidates in the DR4  \citep{2017MNRAS.472.1129P, Petrillo_2018, Petrillo_2019, He_2020, Li_2020}. 

\section{Methodology}
\subsection{Self-attention and Transformer encoder}
\begin{figure}[ht]
\centering
\includegraphics[width=470 pt,keepaspectratio]{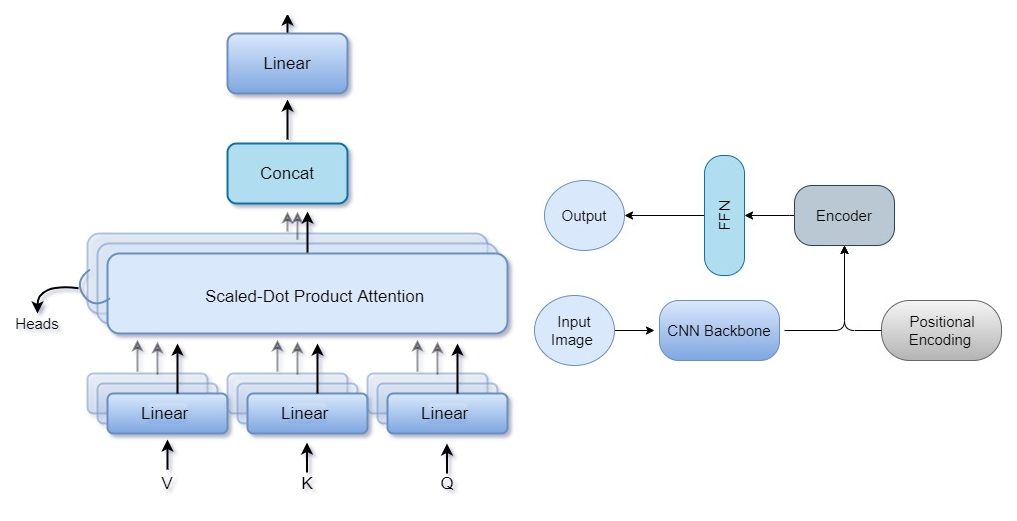}
\caption{Scheme of the multi-head attention layer and scheme of the architecture of the transformer encoder}
\label{fig:Transformer}
\end{figure}

The Transformer models we propose to detect SLs are inspired by the DEtection TRansformer (DETR) created by Facebook, based on the principle of self-attention \citep{carion2020endtoend}. 
In general, the attention function can be defined mathematically as 
$ \text{Attention}(Q,K,V) = \text{softmax} \left( QK^T/\sqrt{d_k} \right) V,$
where $Q, K, V$ are vectors and $\sqrt{d_k}$ is the dimension of the vector key $K$. The \text{softmax} function, by definition, is a normalized exponential function that takes an input vector of K real numbers and normalizes it into a probability distribution proportional to the exponential of the input numbers.
As we compute the normalized dot product between the query ($Q$) and the key ($K$), we get a tensor ($QK^T$) that encodes the relative importance of the features in the key to the query \citep{vaswani2017attention}. For self-attention, the vectors $Q$, $V$, and $K$ are identical. Hence, multiplying the tensor ($QK^T$) by the vector ($V$) results in a vector that encodes the relative importance of features inside the input vector. In simple terms, the central idea of self-attention is to assign relative importance to the features of the input based on the input itself. As shown in Fig. \ref{fig:Transformer}, the transformer encoder has a very simple architecture. The architecture consists of a CNN backbone followed by a set of attention layers that altogether extract and weigh the relevant features of an input. In the final layers, there is a feed-forward network (FFN) that learns how to combine these features to predict the output. 
 
\begin{figure}[ht]
\centering
\includegraphics[width=500 pt,keepaspectratio]{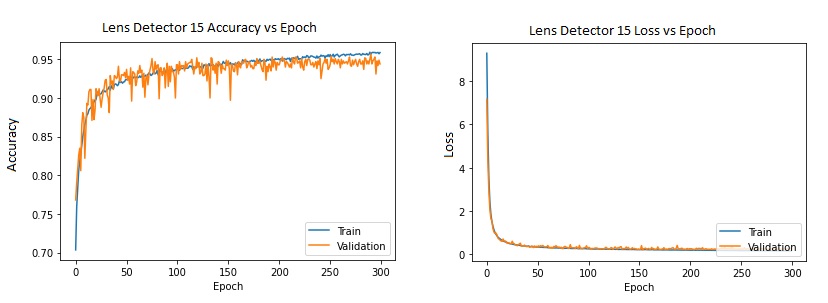}
\caption{Accuracy and loss of the model 'Lens Detector 15' as a function of the epoch during training.}
\label{fig:training}
\end{figure}

 The model we used for our study is the model named 'Lens Detector 15' mentioned in the \citet{Hareesh}. The model  was first trained for 300 epochs with an initial learning rate of $\alpha = 10^{-4}$ to reduce the binary cross entropy loss and again trained for another 100 epochs starting with a learning rate of $\alpha = 10^{-5}$. We use the exponential linear unit (ELU) function as the activation function for all the layers in the model. The weights of the model are initialized with the  Xavier uniform initializer, and all layers are trained from scratch by the ADAM optimizer with the default exponential decay rates \citep{Glorot2010UnderstandingTD,kingma2017adam}. The accuracy and the loss of the model as a function of epochs are plotted in Fig. \ref{fig:training}. For a detailed description of the models, their architecture, and how the hyper-parameters affect the performance, please refer to \citet{Hareesh}. In the spirit of reproducible research, our code for Lens Detector 15 is publicly available at \hyperlink{https://github.com/hareesht23/Lens-Detector}{https://github.com/hareesht23/Lens-Detector}.

\subsection{Visual Inspection}
One of the big differences between simulated data and real data is the presence of labels. Real data do not come with labels, and therefore in order to check the accuracy of a model (how many SL candidates detected by the model are actually lenses), a visual inspection by experts is needed. We trained the model to distinguish lenses from non-lenses. More specifically, lenses are labelled with 1 and other objects with 0. Therefore, when an object is presented to the model, the latter gives a prediction probability of being a lens. For this study, we use a threshold of 0.8; if an object has a prediction probability greater than the threshold, it is considered a lens candidate and further inspected.
In our case, four volunteers visually inspected the strong lens candidates with the task of tagging real lenses from false positives. However, visual inspections can be biased depending on different factors, for this reason, multiple people and a scale of grades are needed in the process. We give a score to the images as follows:
\begin{itemize}
\itemsep-0.75em 
    \item 5 - a sure lens, clear arcs-like structures
    \item 3 - maybe a lens, arcs-like structure but not so resolved
    \item 1 - interesting candidate but most likely not a lens
    \item 0 - not a lens
\end{itemize}
Finally, candidates graded with at least one 5 or identified by at least two people (total grade bigger than 4) were further considered and inspected.

\section{\label{sec:results}Results} 

\subsection{Results on the simulated dataset}
\begin{figure}[ht]
\centering
\includegraphics[width=\textwidth,keepaspectratio]{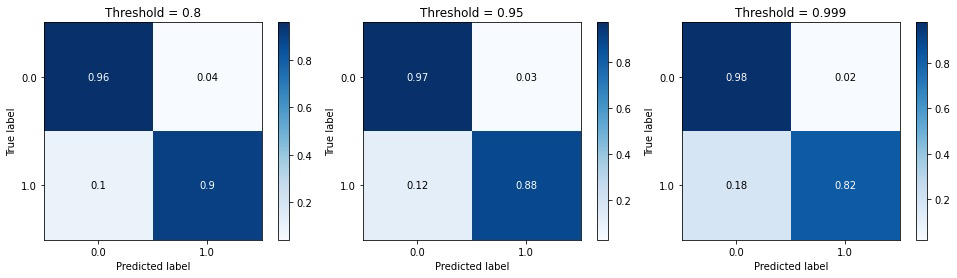}
\caption{The area under the receiver operating characteristic curve (AUROC) and the probability output for the lens detector 15 on the challenge set.}
\label{fig:CM}
\end{figure}

\begin{figure}[ht]
\centering
\includegraphics[width=\textwidth,keepaspectratio]{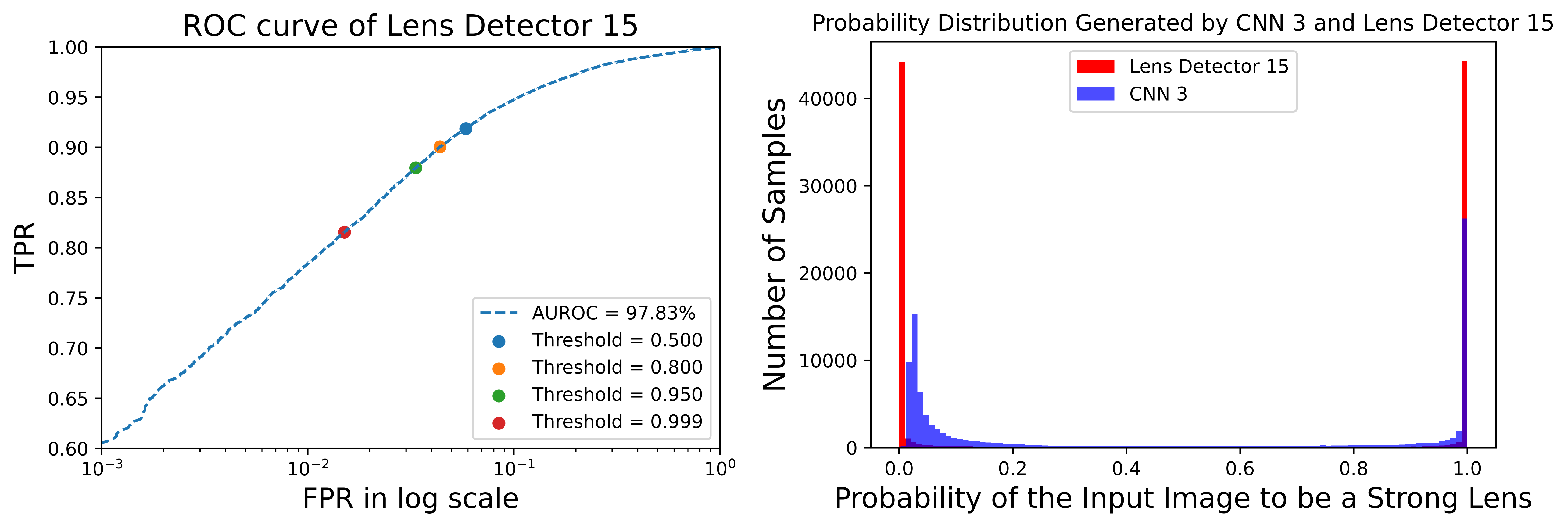}
\caption{The area under the receiver operating characteristic curve (AUROC) and the probability output for the lens detector 15 on the challenge set.}
\label{fig:results}
\end{figure}

The Bologna Lens Challenge was intended to improve the efficiency and biases of tools used to find strong gravitational lenses on galactic scales. During the challenge, it was proven that automated machine learning methods are more efficient in detecting SLs than traditional machine learning methods. Here we present the results from the best self-attention-based encoder 'Lens Detector 15' \citep{Hareesh} that was able to surpass all the other methods that participated in the challenge and the model used in this study on the KiDS survey. The confusion matrices for the challenge dataset with three different thresholds (0.8,0.95, and 0.999, respectively) are shown in Fig. \ref{fig:CM}. It should be noted that even with a very high threshold, such as 0.999, the model is able to identify more than 80\% of the strong lenses. The area under the receiver operating characteristic curve (AUROC) and the probability output for the lens detector 15 on the challenge set are shown in Fig. \ref{fig:results}. For a detailed discussion and review of different encoder models created for the challenge, please refer to \citet{Hareesh}.

\subsection{Results on KiDS DR4}

\begin{figure}[ht]
\centering
\includegraphics[width=0.5\textwidth,keepaspectratio]{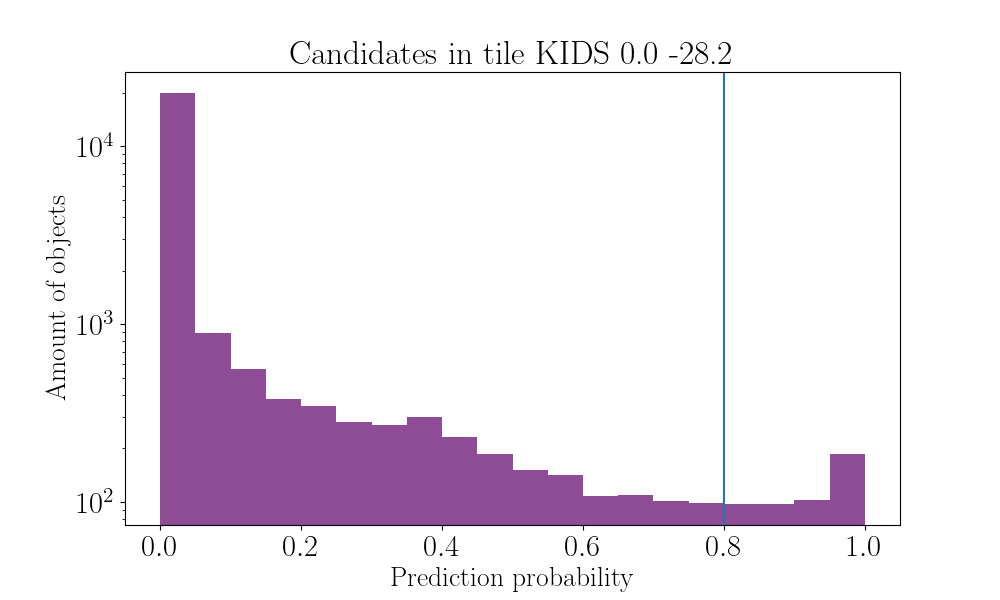}
\caption{The output probability distribution of our model for the KiDS tile $0.0-28.2$. The blue vertical line indicates the prediction probability threshold.}
\label{fig:hist_model_15}
\end{figure}

\begin{figure}[ht]
\centering
\includegraphics[width=\textwidth,keepaspectratio]{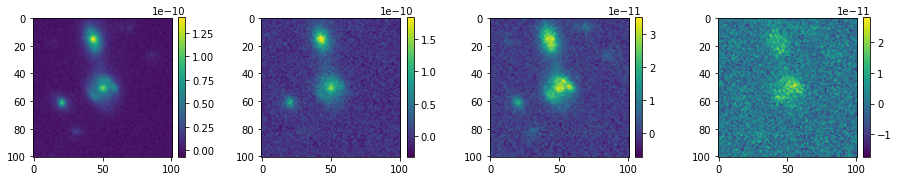}
\caption{A new SL candidate found in this work with a lens redshift = 0.31. (KiDS tile = 0.0,-28.2, Ra= 0.06, Dec = -28.193, the information about the candidate is taken from KiDS DR4.}, 
\label{fig:lens_us}
\end{figure}

The promising results on the simulated dataset encouraged us to test the model on the real KiDS DR4 data. However, since the simulated dataset does not exactly mimic the real survey, the number of false positives (FP) is too high to be eventually useful. For a threshold prediction probability = 0.8, the lenses detected by the model are around 1000 per tile. After a visual inspection, only a few SL candidates remain - the model output is contaminated by FPs. For instance, the  KiDS $0.0-28.2$  tile contains $\approx$25000 galaxies with $z_L<0.8$, of which ~ 900 are detected as SL, but only 5 SL candidates are high-quality candidates selected by the labellers. This gives a rate of 0.5\%  True Positives. The prediction probability distribution for the discussed tile is shown in Fig.\ref{fig:hist_model_15}. 
The model is able to filter out most of the not lensed galaxies, but this result is not good enough for an application on the entire survey; having ~ 1000 candidates per tile means a visual inspection of 10$^5$ objects. One example of a lens found in this first blind search is shown in Fig. \ref{fig:lens_us}, the model is able to identify arc-like structures, but at the same time, it is contaminated with false positives. In the next section, we discuss how to improve this.

\section{Discussion}
An initial glance at the results shows that our model is able to detect strong lens candidates along with a large fraction of false positives. As we explained earlier, all the SL candidates identified by the model should be inspected by human inspectors, which is a time-consuming procedure. For this reason, reducing the number of false positives is a very important step. Hence a detailed interpretation of the results is necessary to understand how to increase the performance of the model when moving from simulated data to real data. 

Firstly, as mentioned above, we did not apply any selection criteria, in the catalogue space, to the candidates (e.g. image quality,  flux density, and/or morphology). This was done to understand whether object preselection impacts the performance. The existing literature on KiDS survey choose a sample of luminous red galaxies (LRGs) or bright galaxies (BGs) because the probability of finding an SLs in this sample is very high \citep{2017MNRAS.472.1129P, Petrillo_2018, Petrillo_2019, He_2020, Li_2020}. This kind of preselection lowers the number of false positives simply because it decreases the objects given, in total, to the model. However, we are not planning to apply this kind of selection to be sure of not losing important SL candidates that the previous searches missed. Finally, our results are in favour of preselection by the image quality; during the visual inspection, we found that the model gives a high probability of being a lens for some poorly observed images (glitches), which could have been removed with the preselection.  

\begin{figure}[ht]
\centering
\includegraphics[width=\textwidth,keepaspectratio]{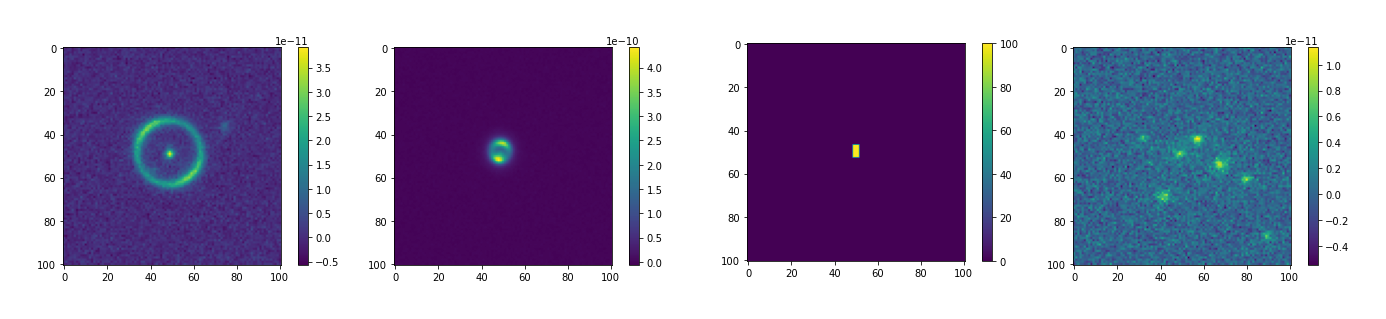}
\caption{Example of images labelled as a strong lens. Each image corresponds to a different strong lens configuration, and the images are only plotted only in the r-band.}
\label{fig:lens_train_r}
\end{figure}

To clearly understand the nature of the False positives, we looked deeper into the training data by visually inspecting the strong lenses in it. Some peculiar examples are shown in Fig. \ref{fig:lens_train_r}; for this purpose, we only show the r-band image, which is the one with the highest resolution. From the figure, it is clear that the first two cases have visible geometrical features of an SL, whereas the last two cases (starting from the left) do not have any evident features. The classification of glitched images as lenses can be explained by the presence of objects like the third lens in Fig. \ref{fig:lens_train_r} in the training data. The central pixel is clearly saturated - resembling a glitch in the charge-coupled device (CCD) camera. Moreover, some point-like objects have been classified as lenses, which is explained by the presence of objects like the fourth lens in Fig. \ref{fig:lens_train_r}.

\begin{figure}[ht]
\centering
\includegraphics[width=\textwidth,keepaspectratio]{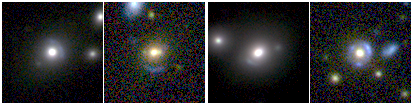}
\caption{Strong Lenses found by \citet{Petrillo_2019}}
\label{fig:lens_them}
\end{figure}

This brings us back to our initial problem of switching from simulated data to real data without having a significant number of false positives (FP). Even though the model is trained on the simulated data to achieve very high accuracy or AUROC, it gets confused when it is shown real data. Earlier works also faced the same problem of going from simulated to real data. For example, among the SLs candidates identified by \citet{Li_2020} using a CNN, only 1.7\% passed the visual inspection. Hence this problem poses an interesting question to the astrophysics community. We plan to tackle this problem using transfer learning and other data-augmenting methods. Instead of training the model with more simulations, we train the existing model on the set of 169 SL candidates found by the KiDS collaboration (examples in Fig.\ref{fig:lens_them}) and use data augmentation to increase the sample size. For the non-lens class, we plan to use the false negatives uncovered by the model. For this class, we do not need data augmentation and to ensure the dataset is not biased toward one part of the sky we will use the same amount of objects per tile with the highest prediction probability of being a lens. Giving the 'most' wrong examples should help with the model performance. The training set will be balanced, with the same amount of real (augmented) lenses and non-lenses. \\
Fine-tuning the model parameters with transfer learning on small datasets can improve the model performance, which will tackle the issue of the lack of large realistic datasets. We have verified this idea and found that it is a promising approach, and the final results will be discussed in further publications.

\section{Conclusions}
In \citep{Hareesh}, we have proposed a novel machine learning approach, built on self-attention-based encoders, to detect strong gravitational lenses. In this follow-up study, we have explored the usage of this architecture on the KiDS survey and investigated the pitfalls of applying a simulation-trained model on real data. The poor performance of the model on survey data is demonstrated by the inability of the simulated events to capture the high complexity of real data and the presence of mislabelled events.  
For very rare events such as SLs, it is challenging to have a labelled dataset, and it is imperative to look for new methods for training models rather than sticking with simulations only. We propose transfer learning and data augmentation as practical solutions to this problem, and we plan to discuss them further in the upcoming publications.

\begin{acknowledgments}
We would like to acknowledge the support of dr. Orest Dorosh and Marianna Zadrożna for helping with the visual inspection. Authors and NCBJ are grateful for financial support from MNiSW grant DIR/WK/2018/12, and NCN grants UMO-2017/26/M/ST9/00978 and UMO-2018/30/M/ST9/00757.
\end{acknowledgments}

\bibliography{manuscript}

\end{document}